# Quantum Imaging with Undetected Photons

Gabriela B. Lemos,[1,2] Victoria Borish,[1,3] Garrett D. Cole,[2,3] Sven Ramelow,[1,3]

Radek Lapkiewicz,[1,3] and Anton Zeilinger[1,2,3]

[1]*Institute for Quantum Optics and Quantum Information, Boltzmanngasse 3, Vienna A-1090, Austria* [2]*Vienna Center for Quantum Science and Technology (VCQ), Faculty of Physics, University of Vienna, A-1090 Vienna, Austria* [3]*Quantum Optics, Quantum Nanophysics, Quantum Information, University of Vienna, Boltzmanngasse 5, Vienna A-1090, Austria*

## SUMMARY

Indistinguishable quantum states interfere, but the mere possibility of obtaining information that could distinguish between overlapping states inhibits quantum interference. Quantum interference imaging can outperform classical imaging or even have entirely new features. Here, we introduce and experimentally demonstrate a quantum imaging concept that relies on the indistinguishability of the possible sources of a photon that remains undetected. Our experiment uses pair creation in two separate down-conversion crystals. While the photons passing through the object are never detected, we obtain images exclusively with the sister photons that do not interact with the object. Therefore the object to be imaged can be either opaque or invisible to the detected photons. Moreover, our technique allows the probe wavelength to be chosen in a range for which suitable sources and/or detectors are unavailable. Our experiment is a prototype in quantum information where knowledge can be extracted by and about a photon that is never detected.

I. INTRODUCTION

Information is essential to quantum mechanics. In particular, quantum interference occurs if and only if there exists no information that allows one to distinguish between the interfering states. It is not relevant if an observer chooses to notice this information or not. These are not just conceptual issues; they have direct practical consequences. In fact, interference of a photon can reveal information of another photon, which is not detected. As an example, here we apply this as a method of quantum imaging[1].

In imaging, the ideal wavelength for illuminating the object normally depends on both the properties of the object to be imaged and the wavelength sensitivity of available detectors. This makes low-light imaging very difficult at wavelengths outside the range for which low-light cameras are available. In order to circumvent this problem, one approach has used optical non-linearity to convert the light coming from the object into a shorter wavelength where efficient and low-noise detectors are available[2,3]. Another method is two-colour ghost imaging[4], in which the light field of one wavelength is used to illuminate an object and the image appears in the correlations between this and a light field in another wavelength, thus requiring coincident photon detection at both wavelengths[5].

Our quantum imaging technique is based on a quantum interference phenomenon that was first shown in the early 1990's[6]. We begin with an explanation of this experiment, which is illustrated in Fig. 1. A pump laser beam divides at a 50:50 beam splitter (BS1) and coherently illuminates two identical non-linear crystals, NL1 and NL2, where pairs of collinear photons called signal (yellow) and idler (red) can be created. The probability that a down-conversion occurs at each crystal is equal and very low so the chance that more than one pair of photons is produced at a given time can be neglected.

The idler photons created in NL1 are reflected at the dichroic mirror D1 into spatial mode $d$, and signal photons pass into spatial mode $c$. The idler then passes through the object $O$ that has a real transmittance coefficient $T$ and imparts a phase shift $\gamma$. We write this as $|c\rangle_s|d\rangle_i \to Te^{i\gamma}|c\rangle_s|d\rangle_i + \sqrt{1-T^2}|c\rangle_s|w\rangle_i$ where, for simplicity, we assume that all the idler photons that are not transmitted occupy a single state $|w\rangle_i$. After being reflected at dichroic mirror D2, the idler photons from NL1 are perfectly aligned with idler photons produced at NL2, $|d\rangle_i \to |f\rangle_i$. The quantum state at the grey dotted line in Fig.1 can be written as

$$\frac{1}{\sqrt{2}}[(Te^{i\gamma}|c\rangle_s + |e\rangle_s)|f\rangle_i + \sqrt{1-T^2}|c\rangle_s|w\rangle_i]. \qquad (1)$$

The idlers are now reflected at dichroic mirror D3 and are not detected. The signal photon states $|c\rangle_s$ and $|e\rangle_s$ are combined at the 50:50 beam splitter BS2. The detection probabilities at the outputs, $|g\rangle_s$ and $|h\rangle_s$, are obtained by ignoring (tracing out) the idler modes, giving

$$P_{g/h} = \frac{1}{2}[1 \pm T\cos\gamma]. \qquad (2)$$

This formula shows that fringes with visibility T can be seen at either output of BS2, even though the signal beams combined at BS2 have different sources. These fringes appear in the signal single photon counts; no coincidence detection is required. This is possible because the idler photon that is reflected at the dichroic mirror D3 does not carry any information about the crystal where it was created, and therefore the two modes of the signal interfere when overlapped by BS2, *i.e.* now each signal photon came from *both* NL1 *and* NL2.

A very peculiar feature of this interferometer is that no photon that reaches the detectors can have gone through path $d$ (Fig. 1). Yet, in our experiment, it is precisely here that we put the object we want to image. The key to this experiment is that the signal source information carried by the undetected idler photons depends on T. For instance, if T = 0, one could monitor the idlers reflected at D3. If an idler photon and a signal photon at $|g\rangle_s$ or $|h\rangle_s$ were detected in coincidence, the observer would know that the signal photon was produced in NL2. The detection of a signal photon *without* the detection of a corresponding idler photon would imply that the signal photon was produced in NL1. This *which-source* information destroys interference because it makes the quantum states overlapping at BS2 distinguishable. If T =1, the idler photons do not carry any information about the source of the detected signal photon. The signal states at the output of BS2 are then indistinguishable, thus the interference term in Eq. 2 appears. It is important to emphasize that one does not actually have to detect the idler photons, for it is only the *possibility* of obtaining which-source information that matters in this experiment.

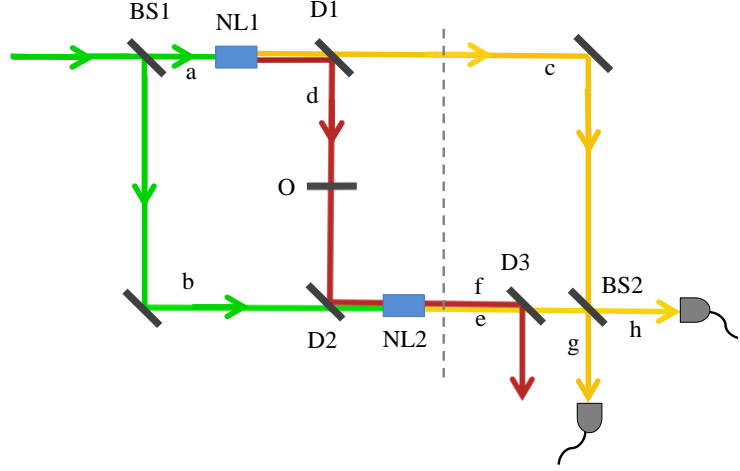

**Figure 1. Schematic of the experiment.** Laser light (green) splits at beam splitter BS1. Beam *a* pumps nonlinear crystal NL1 where collinear down-conversion may produce a pair of photons of different wavelength called signal (yellow) and idler (red). The idler, bearing amplitude and phase information from object O, reflects at dichroic mirror D2 to align with pump mode b, which illuminates nonlinear crystal NL2, whose own collinear idler aligns with the NL1 idler to make the down-conversion source unknowable, as the two signals combine at beam splitter BS2. Consequently, signal interference at BS2 reveals idler transmission properties of object O.

Our experiment has a connection to interaction free measurements[7,8]. Notice that $P_h = 0$ if no object is placed in the setup ($T = 1$ and $\gamma = 0$). Now consider that an opaque object ($T = 0$) is placed in mode $d$ and we detect the idlers reflected from D3. In this case, $P_h > 0$, and so a click in $|h\rangle_s$ with a corresponding click in the idler detector would indicate that an object is present even though no detected photon interacted with the object. With our experimental setup it is thus possible to realize interaction free imaging of an object.

Several steps change the Fig. 1 arrangement into an imaging arrangement (Fig. 2). We replace the uniform object with one bearing features, *i.e.* $T$ and $\gamma$ depend on transverse position, and the photon counters are replaced with cameras. The signal and idler photon pairs are produced with sharp transverse spatial correlations[9,10], and lenses image the object surface onto the camera[11,12]. The image associated with a non-constant transmittance is due to local *which-source* information carried by the undetected idler photons. The image of a phase distribution is of an entirely different nature: it is due to the fact that the position dependent phase shift imprinted on the idler photons in path $d$ is actually passed to the signal; *i.e.*, $|c\rangle_s(Te^{i\gamma}|d\rangle_i) = Te^{i\gamma}|c\rangle_s|d\rangle_i$.[13] In this case, the idler beam $|f\rangle_i$ does not even carry the phase pattern and by itself (without also the detection of the sister signal photon in coincidence) could not be used to obtain the phase image[14,15].

With O and D2 removed, equation (1) would be an ordinary two-particle entanglement[13], $i|c\rangle_s|d\rangle_i - |e\rangle_s|f\rangle_i$. With them in, $|d\rangle_i \rightarrow iTe^{i\gamma}|f\rangle_i$, creating (1). A normal two-particle entanglement has blossomed into an exciting single-particle superposition, especially rich when T and γ are transverse-position dependent.

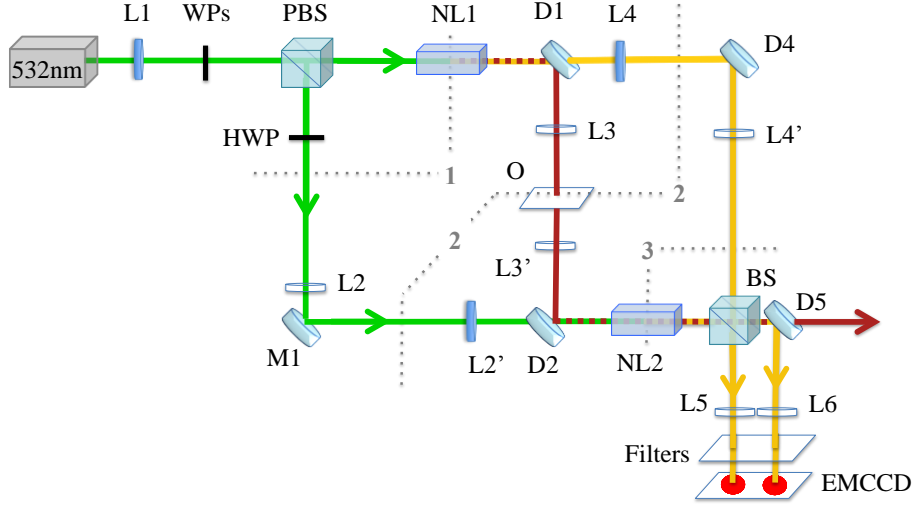

**Figure 2: Experimental setup.** A continuous 532 nm laser (green) illuminates crystals NL1 and NL2. Wave plates (WPs) adjust the relative phase and intensity of the outputs of the polarising beam splitter (PBS). The dichroic mirror D1 separates 810 nm down-converted photons (yellow) and 1550 nm photons (red). The 1550 nm photons are transmitted through the object O and sent through NL2 by dichroic mirror D2. Lenses image plane 1 onto plane 3, and plane 2 onto the EMCCD camera. A 50:50 beam splitter (BS) combines the 810 nm beams. Dichroic mirrors D1, D4, and D5 transmit the pump.

## II. EXPERIMENT

A detailed schematic of our imaging setup is shown in Fig. 2. A 532 nm linearly polarised Gaussian pump laser beam focused by lens L1 on plane 1, is divided at a polarizing beam splitter (PBS) and then coherently illuminates two identical periodically poled potassium titanyl phosphate (ppKTP) crystals, NL1 and NL2. The role of the PBS is to act together with wave plates (WPs) as a tuneable beam splitter, such that we can control the relative amplitudes and relative phases between the reflected and transmitted pump beams. An extra half wave plate (HWP) rotates the polarisation of the reflected pump beam just after the PBS, such that both beams have the same linear polarization. If down-conversion occurs in NL1, the 1550 nm idler is reflected by dichroic mirror D1. Both the 810 nm signal and the pump are transmitted through D1 and are separated from one another at dichroic mirror D4, which transmits 532 nm light and reflects 810 nm light. A long-pass filter (not shown in the figure) placed directly before the object O prevents any residual 532 nm or 810 nm light reaching

NL2. The 1550 nm beam that originated in NL1 illuminates the object O and is then overlapped with the pump beam at dichroic mirror D2 that transmits 532 nm light and reflects 1550 nm light.

Lens pairs L2 - L2', L3 - L3', and L4 - L4' image plane 1 onto plane 3, thereby ensuring that pump, idler and signal, respectively, are identical in these planes. Finally, lenses L5 and L6 together with L2', L3', and L4' image object plane 2 onto the camera surface.

The detection of the 810 nm photons is realised in both outputs of the beam splitter using an Electron Multiplying Charge Coupled Device (EMCCD) camera. This camera is capable of single-photon sensitivity for illumination at 810 nm but has a negligible response at 1550 nm. Nonetheless, a combination of spectral filters is used before the camera to ensure that neither 1550 nm photons nor 532 nm pump photons reach the camera. The detected 810 nm photons have a spectral bandwidth of 3 nm. They are detected with no heralding. See the Methods section for additional discussion about wavelength filtering, details of the imaging system, and important optical path length differences.

## III. RESULTS

In this section we show images produced by three different objects. In the first experiment, a cardboard cut-out is placed into the path D1-D2. Due to transverse spatial correlation between the signal and idler photons[16], interference is only observed in the region of the 810 nm output beams corresponding to 1550 nm photons transmitted through the shape cut out of the cardboard. The region corresponding to the blocked 1550 nm photons presents no interference (equal intensity at each output of the beam splitter) as the cardboard in this region (T=0) acts as a detector that *could* be used to obtain position dependent *which-source* information. The next two objects are used to demonstrate that our setup can image a position dependent phase shift. We first obtain the image of an etched silicon plate that is opaque to 810 nm light but transparent to 1550 nm light. Then we image an etched fused silica plate that is invisible when placed in the (detected) 810 nm beam but it becomes visible when placed in the undetected 1550 nm beam.

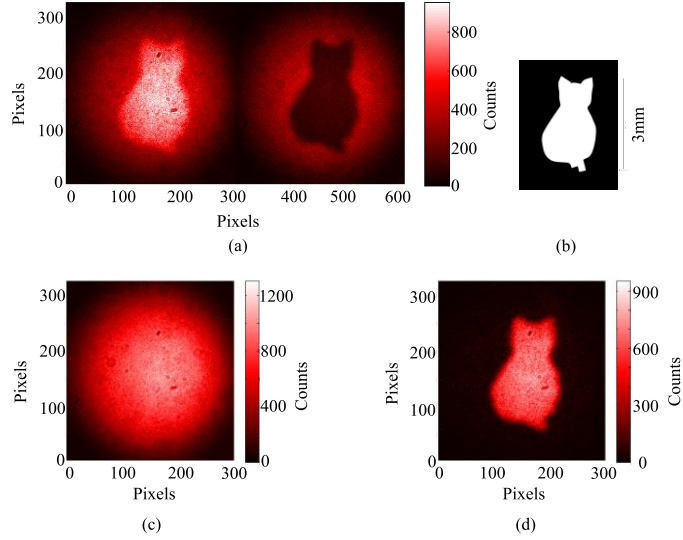

**Figure 3: Intensity Imaging. a,** constructive and destructive interference at the output of BS when we placed in path D1-D2 the cardboard cut-out shown in **b**. The scales show the count range per pixel (16 x16 μm) in an exposure time of 0.5 seconds with an electron multiplying gain factor of 20. The pump power was 150 mW. **c,** the sum of the outputs gives the intensity profile of the signal beams. **d,** the subtraction of the outputs leads to an enhancement of the interference contrast, as it exhibits the difference of constructive and destructive interference.

Fig. 3a shows the output of BS when a cardboard cut-out (illustrated in Fig. 3b) is inserted in the path D1-D2. Constructive interference is seen at one output of the beam splitter and destructive interference is observed in the other output. To clearly demonstrate that interference only occurs in the region corresponding to the idler beam transmitted through the shape cut out of the cardboard, we show in Figs. 3c and 3d respectively the sum and difference of the complementary images. The sum of the two outputs of the BS gives the featureless intensity profile of the signal beams. This shows that the signal beams, while carrying the cut-out information, are not absorbed at all by the mask. The subtraction of the intensity pictures exhibits the difference of constructive and destructive interference, and the background cancels out.

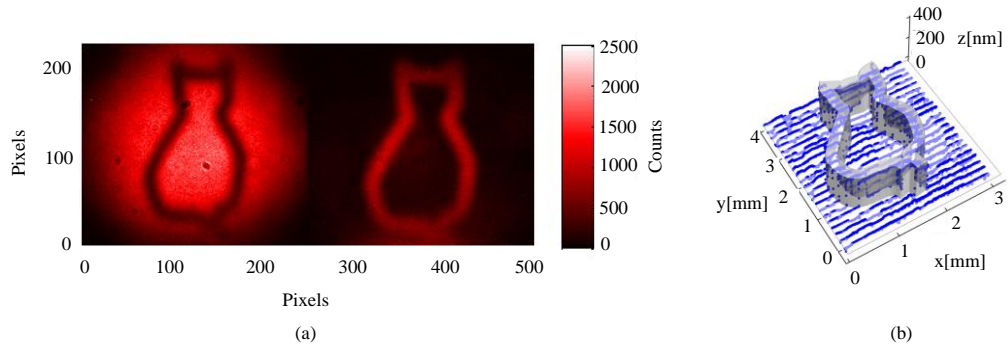

**Figure 4. Phase image of an object opaque for 810 nm light. a,** detection of 810 nm photons at both outputs of BS when a silicon plate (opaque to 810 nm light) with a 3 mm tall etched cat was introduced in path D1 - D2. The scale shows counts per pixel (16 x16 µm) in an exposure time of 0.5 seconds with an electron multiplying gain factor of 20. The power of the pump beam was 150 mW. **b,** 3D rendering of the etch design overlaid with scans (blue points) of the actual etch depth.

In Fig. 4a, we show the image of a 500 $\mu$m thick silicon plate etched with the shape shown in Fig. 4b (see Methods section for details of the silicon plate and the etching process). Silicon is opaque for illumination at 810 nm but highly transparent at 1550 nm. Thus it is not possible to illuminate the silicon with 810 nm light and obtain a transmission image. However, when we place the object in path D1-D2, the difference in optical path length for the parts of the 1550 nm beam going through the etched and non-etched regions imparts a relative phase shift of π, which can be seen in the detected 810 nm at the output of BS2 (Fig. 4a). Even though our camera is blind to 1550 nm light, we can realize an image with our setup because different wavelengths are used for illumination and detection.

Finally, Fig. 5a shows the image of an etched fused silica ($SiO_2$) plate (details are in the Methods section). We take advantage of the flexible phase matching conditions in our ppKTP crystals and lower the temperature in order to obtain collinear non-degenerate down-conversion at 820 nm and 1515 nm. The object showing the Greek letter ψ (Fig. 5b) has an etch depth of 1803 nm, which imparts a relative phase shift of approximately 2π for 820 nm light. Thus the object is invisible when placed in between L4 and L4' (top of Fig. 5a). This same etch depth gives a phase shift of approximately π for 1515 nm light, so when this same object is placed in the path D1-D2, an image seen in the contrast of constructive to destructive interference is retrieved in the 820 nm output (bottom picture of Fig. 5a).

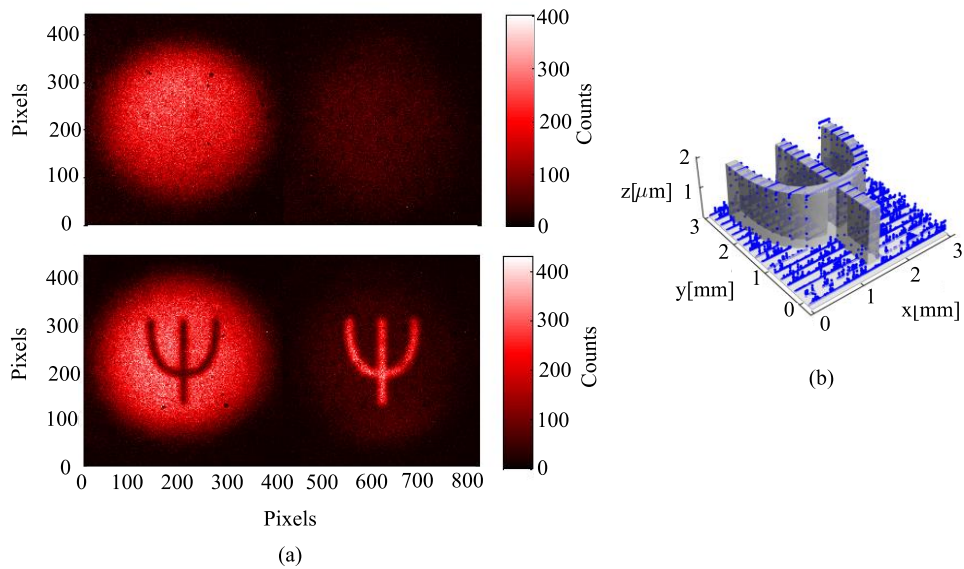

(a)

**Figure 5. Phase imaging of a 2π step at 820 nm**. **a,** the top picture was taken with the object placed in the 820 nm beam in between L4 and L4', and in the bottom picture the object was placed in the 1515 nm beam in path D1-D2. The scale shows counts per pixel (16 x16 μm) in an exposure time of 0.5 seconds with an electron multiplying gain factor of 20. The power of the pump beam was 150 mW. **b,** 3D rendering of the design overlaid with scans (blue dots) of the actual etch depth.

In order to quantify the visibility in our imaging experiment, we detect the total intensity of 810 nm photons at one output of BS as a function of the relative phase between the pump beams that illuminate each crystal. Fig. 6 shows a plot of the count rate measured with an avalanche photodiode when no object is present. The red circles show the experimental points, and the best fitting sinusoidal function (red line) gives a visibility of (77 ± 1)%. This is much higher than the visibility of ~30% obtained by Zou, *et al*[17]. The visibility for our experiment is given not only by losses in both the 1550 nm and 810 nm arms of the interferometer, but also by our imperfect alignment technique for the two idler beams. The blue squares correspond to data obtained when the path NL1-NL2 is completely blocked, which results in zero interference visibility. Interference only arises if the idler between the two crystals is unblocked, for only then is its source, and therefore also the source of its signal sister, unknowable.

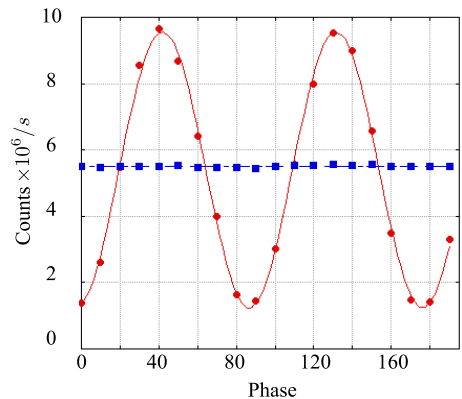

**Figure 6. Visibility of the experiment.** The count rates were recorded with the path D1 to D2 both unblocked (red dots) and blocked (blue squares) as the relative phase between the transmitted and reflected beams of the PBS was varied. The red line is a sine curve fit for the experimental data giving $(77 \pm 1)\%$ visibility. The error bars are smaller than the size of the data points.

IV. DISCUSSION

We present here a new quantum imaging system for intensity and phase imaging where the photons that illuminate the object are never detected. This allows us to illuminate a object with a wavelength that is difficult to detect, but detect only photons of a more convenient wavelength, one for which the object was either opaque or invisible. This experiment is fundamentally different to ghost imaging[4,5], because it does not require coincidence detection. Furthermore, our technique could be used for non-degenerate interaction-free imaging.

Our system can easily be adjusted to realize grey scale intensity or phase imaging, and it can be adjusted to measure spectral features (spectral imaging)[18,19]. This kind of imaging, particularly in mid- or near-infrared, is an increasingly relevant technology with a wide variety of potential applications in environmental studies[20], medical diagnostics[21], and other areas[2,22,23].

Our technique does not require the laser or the detector to function at the same wavelength as that of the light probing the object. Additionally, the use of down-conversion as the source provides flexibility in the wavelengths for both detection and illumination of the object. Indeed, an object can be probed with light ranging from UV through MIR or possibly even the THz regime while the image is detected at a freely chosen wavelength where detectors are technologically available or exhibit superior performance.

We have shown that information can be extracted about a photon without detecting it. Knowing the two-photon state one can obtain the information about an object. It has not escaped our attention that on the other hand, knowing the object one could obtain information about the quantum state without detecting it.

## METHODS

### A. Down-conversion Sources

The 532 nm pump beam is generated by a Coherent Sapphire SF laser and is slightly focused onto the two periodically poled potassium titanyl phosphate (ppKTP) crystals with dimension $1 \times 2 \times 2$ mm$^3$ and poling period 9.675 µm for type-0 phase matching. The crystals are spatially oriented so the down-conversion occurs when the cw pump beam is horizontally polarised (both the signal and idler produced are also horizontally polarised). In order to conform to the phase-matching conditions for 810 nm and 1550 nm photons and to ensure that the spectra of the down-conversion are precisely centred on the same wavelength for both crystals, NL1 (NL2) is heated to 83.7 °C (84.7 °C). When the set-up is adjusted to produce 820 nm and 1515 nm photons (to be used with the fused silica phase object), NL1 (NL2) is heated to 39.2 °C (39.7 °C).

### B. Wavelength Filtering

Inside the interferometer, D1 is used to separate the 810 nm photons from the 1550 nm photons. Mirror D1 (and also D2) reflect about 93% of 1550 nm light and transmit about 99% of 810 nm light. Most of the pump beam going through NL1 is transmitted through both D1 and D4 (each with a transmittance at 532 nm of around 97%) and therefore never reaches BS. The dichroic mirror D5 additionally transmits some 532 light (around 25%), so some of the pump beam that goes through NL2 as well as some of the remaining pump beam from NL1 are discarded there. All remaining pump beam light is cut out with either filters or the object. The silicon object is opaque to both 532 nm and 810 nm light, thus blocking these wavelengths along the path D1-D2. When the other objects are used, a long-pass filter (Semrock) is placed just before the object to cut out these lower wavelengths. The remaining 532 nm light that is not separated out through the dichroic mirrors or object is blocked in front of the camera by three filters. A 3 nm narrowband filter centred at 810 nm and two long pass filters (Semrock) were attached directly to the front of the camera. As it utilizes a silicon-based detector, the Andor Luca-R EMCCD camera does not detect 1550 nm photons.

### C. Imaging Lens Systems

As it is crucial that the down-converted photons be identical, we use confocal lens systems to image plane 1 onto plane 3 (see Fig. 2), thus ensuring that the pump beams at NL1 and NL2 are identical, the 810 photons when they combine at the BS are identical, and the 1550 nm photons are identical from NL2 onward. Lenses L2 and L2' image plane 1 of the pump onto plane 3, and similarly L3 (L4) and L3' (L4') image plane 1 onto plane 3 for the 1550 nm (810 nm) photons. Lenses L5 and L6 in combination with L4' image plane 2 onto the EMCCD camera. Lenses L2, L2', L3, L3', L4, L4' have a focal length of $F_1$=75 mm. The distance from plane 1 to each of L2, L3, and L4 is 75 mm; from those lenses to plane 2 is another 75 mm; from plane 2 to L2', L3', and L4' is also 75 mm; and from those lenses to plane 3 is yet another 75 mm. This ensures that the photons produced in both crystals have the same waist and divergence when they reach the BS. Lenses L5 and L6 have a focal length of $F_2$=150 mm. They are placed 150 mm after plane 3 and 150 mm before the camera. The total imaging magnification from the object to the camera is given by $\frac{F_2 \lambda_s}{F_1 \lambda_i}$, where $\lambda_{s/i}$ are the wavelengths of the signal and idler photons respectively.

### D. Optical Path Lengths

In our single photon interferometer the paths D1-D3-BS and D1-D2-BS need to be equal, even though no detected photons actually follow the entire path D1-D2-BS. To assure indistinguishability of the emission in the two crystals: NL1 and NL2,, the time delay between the arrival of the signal and idler for each of the two crystals must be the same. Assume both the signal and idler are measured immediately after the BS. The path length difference between the signal and idler for the pair from NL1 is the distance from NL1-D1-D2-BS subtracted from the distance NL1-D1-D3-BS. The path length difference between the signal and idler for the pair from NL2 is zero since the down-conversion is collinear. Thus, we see that the optical path lengths between D1-D3-BS and D1-D2-BS must be equal to within the coherence length of the photons. The coherence length of the photons is limited by the filtering (3 nm), so we approximate the coherence length to be 0.2 mm. The other relevant optical path lengths are the paths PBS-D1-D2-NL2 and PBS-M1-NL2. The difference in distance between these paths must be within the coherence length of the laser, which in our case is approximately 200 meters.

### E. Intensity Object

Our intensity object is constructed from 0.33 mm thick card stock with images defined by laser cutting. The images on the object were each 3 mm high.

### F. Microfabricated Silicon Phase Object

The first custom phase object consists of 500-μm-thick double-side polished (100)-oriented single-crystal silicon with imaging targets defined on one face using standard microfabrication techniques. The absorption coefficient of silicon is ~1000 $cm^{-1}$ at 810 nm[24], and it is ~$10^{-4}$ $cm^{-1}$ at 1550 nm[25]. Processing begins by cleaving a 75-mm diameter silicon wafer to obtain chips with lateral dimensions of 25 × 25 mm. The cleaved chips are patterned using conventional optical contact lithography followed by plasma etching. In order to generate a π-phase shift at 1550 nm, features are etched to a depth of approximately 310 nm (nominal height of 321 nm using a refractive index of silicon of 3.48[26]) into the exposed Si surface using a cryogenic (−108 °C) $SF_6/O_2$ reactive-ion etching (RIE) process protected with a positive photoresist mask. To improve thermal transfer, the silicon chips are mounted to a carrier wafer using a thin layer of vacuum grease. Additionally, in order to minimize variations in the overall etch depth and thus resulting phase shift from the imaging targets, the feature linewidth is kept constant over the lithographic pattern to mitigate the effects of aspect-ratio dependent etching (or "RIE lag"). After etching, the chips are removed from the carrier wafer and the masking resist and mounting film are stripped using a combination of organic solvents and oxygen plasma ashing. In order to eliminate spurious reflections from the polished surfaces, a dual-sided silicon nitride anti-reflection (AR) coating is deposited via plasma-enhanced chemical vapour deposition (PECVD) using He-diluted $SiH_4$ and $NH_3$ as reactive process gases. The deposition process yields quarter-wave optical thickness layers at a target film thickness of 2040 Å (with a refractive index of 1.9 at the imaging wavelength of 1550 nm).

In order to achieve the highest contrast, the path length difference between the etched and non-etched regions should be equal to a half wavelength of 1550 nm light adjusted for the difference in the indices of refraction of silicon and air. This gives a target thickness difference of 321 nm (for a refractive index of silicon of $3.48^{26}$). Given the slight error in etch depth; the actual thickness difference is 310 nm, which is still sufficient to obtain high contrast images.

### G. Microfabricated Fused Silica Phase Object

Similar to the silicon phase object described above, the fused silica phase object, cleaved from a 500-µm-thick glass wafer, is constructed via a standard lithographic and reactive ion etching process. In this case the same mask pattern is once again defined with contact lithography. In order to transfer the features into the fused silica, a high-power inductively coupled plasma (ICP) RIE process is required (150 W ICP, 250 W RF powers) with an etch chemistry consisting of $SF_6$ and Ar. Given the poor selectivity to the masking resist, a thick (10 µm) coating of AZP4620 photoresist is required. The target etch depth of 1788 nm is achieved within roughly 10 minutes at room temperature. Given the high plasma energy, thermalization with the cooled carrier wafer is key. Due to non-uniformities in thermal contact with the carrier, we observe significant variation in etch depth (±200 nm) across the surface of the 25 × 25 mm pattern. No AR coating is employed given the small Fresnel reflection (4%) from the low-index silica substrate.

For 820 nm light, an exact 2π phase shift is given by a thickness difference of 1811 nm (using an index of refraction of 1.45)[27]; after processing, the average etch depth recorded for the fused silica sample is 1803 nm.

### H. Showing that stimulated emission is negligible in the experiment

In order to demonstrate in our experiment that the 1550 nm photons from NL1 do not induce down-conversion in NL2, we show in Fig. 7 the count rates for 810 nm photons originating at NL2 when the 1550 nm beam in between D1 and D2 was blocked (blue crosses) and unblocked (red dots). The mean count rate and the standard deviation were obtained by analysing data obtained during 40 seconds. The blue diamonds show that the ratio of the count rates for the blocked and unblocked configuration is very close to 1 irrespective of the pump power.

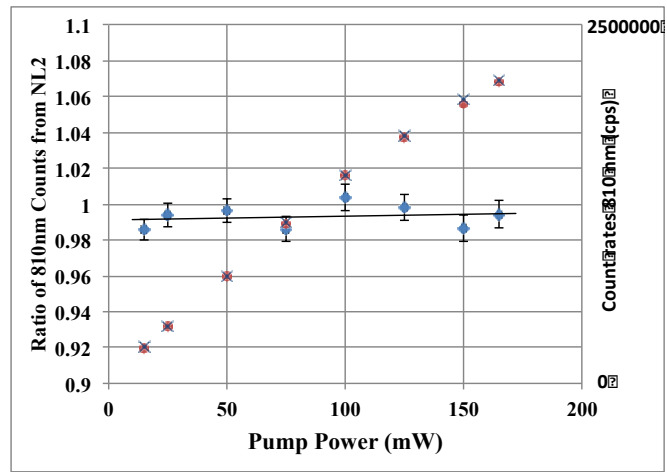

**Fig 1. Showing no induced emission.** The count rates for 810 nm photons produced in NL2 when the path between D1 and D2 was blocked (blue crosses) and unblocked (red dots). The blue diamonds show the ratio of the count rates for the blocked and unblocked configuration. The linear fit for this data (black line) gives an angular coefficient of (2±4) x10$^{-5}$ (mW)$^{-1}$.

**Acknowledgments:** We thank Michael Horne for careful reading of the manuscript, clarifying suggestions and many fruitful discussions. We thank Patricia Enigl for designing the figures for the objects. We thank Daniel Greenberger and Sascha von Egan-Krieger for interesting discussions. We thank Christoph Schaeff for equipment loans. Microfabrication of the phase objects was carried out at the Centre for Micro- and Nanostructures (ZMNS) of the Vienna University of Technology. We gratefully acknowledge Daniela Ristanic for assistance with cryogenic Si etching and Markus Schinnerl for contact object production. GBL is funded by the Austrian Academy of Sciences (ÖAW) through Vienna Center for Science and Technology (VCQ). This project is supported by the Austrian Academy of Sciences (ÖAW), European Research Council (ERC Advanced Grant No. 227844 "QIT4QAD"; SIQS, No. 600645 EU-FP7-ICT), and the Austrian Science Fund (FWF) with SFB F40 (FOQUS) and W1210-2 (CoQus).

Correspondence and requests for materials should be addressed to gabriela.barreto.lemos@univie.ac.at and anton.zeilinger@univie.ac.at.